\documentclass[fleqn,usenatbib]{mnras}

\usepackage[T1]{fontenc}

\DeclareRobustCommand{\VAN}[3]{#2}
\let\VANthebibliography\thebibliography
\def\thebibliography{\DeclareRobustCommand{\VAN}[3]{##3}\VANthebibliography}

\usepackage{graphicx}	
\usepackage{amsmath}	
\usepackage{amssymb}	

\def\gtorder{\mathrel{\raise.3ex\hbox{$>$}\mkern-14mu
             \lower0.6ex\hbox{$\sim$}}}
\def\ltorder{\mathrel{\raise.3ex\hbox{$<$}\mkern-14mu
             \lower0.6ex\hbox{$\sim$}}}

\title[Time Delays]{Measuring time delays: I. Using a flux time series that is a linear combination of time-shifted light curves}

\author[Springer \& Ofek]{Ofer M. Springer,$^{1,2}$
Eran O. Ofek$^{1}$\thanks{E-mail: eran.ofek@weizmann.ac.il}
\\
$^{1}$Department of Particle Physics and Astrophysics, Weizmann Institute of Science, 76100 Rehovot, Israel.\\
$^{2}$Benin School of Computer Science and Engineering, The Hebrew University of Jerusalem, 9190416 Jerusalem, Israel.}

\pubyear{2015}

\begin{document}

\label{firstpage}
\pagerange{\pageref{firstpage}--\pageref{lastpage}}
\maketitle

\begin{abstract}
Several phenomena in astrophysics generate light curves with time delays.
Among these are reverberation mapping,
and lensed quasars.
In some of these systems, the measurement of the time-delay is complicated by the fact that the delayed components are unresolved and that the
light curves are generated from a red-noise process.
We derive the likelihood function of the observations given a model
of either a combination of time-delayed light curves
or a single light curve (i.e., the null hypothesis).
This likelihood function is different from the auto-correlation function investigated by previous studies.
We demonstrate that given a single-band light curve
that is a linear combination of two (or more) time-shifted copies of an original light curve, generated from a red-noise probability distribution,
it is possible to test if the total-flux light curve is a composition of time-delayed copies or, alternatively, is consistent with being a single copy of the original light curve. We also demonstrate that,
in some realistic cases, it is possible to measure the time delays and flux ratios between these unresolved components
even when the flux ratio is about $1/10$.
In the era of synoptic sky surveys, this method is useful for identifying lensed quasars and simultaneously measuring their time delays, and also for estimating the reverberation time scales of active galactic nuclei.
In a companion paper, we build on these results to derive another method that uses the center-of-light astrometric position (e.g., of a lensed quasar)
along with the combined flux. The combined flux and astrometry allow us to identify lensed quasars and supernovae and measure their time delays, with potentially higher fidelity compared to the flux-only method described in the current work.
The astrometry $+$ flux method, however, is not suitable for quasar reverberation mapping. 
We also comment on the commonly used method of fitting a power-law model to a power spectrum,
and present the proper likelihood function for such a fit.
We test the new method on simulations and provide Python and MATLAB implementations.
\end{abstract}


\section{Introduction}
\label{sec:intro}

Several classes of astrophysical objects produce
light curves with time delays, and there is great incentive to measure these time delays.
For example, measuring the reverberation time delay between the continuum and line flux in
Active Galactic Nuclei (AGN) is central for estimating the mass of massive black holes
(e.g., \citealt{Peterson1993_ReverberationMapping}, \citealt{Kaspi+2000_ReverberationQSO}, \citealt{Kaspi+2005_ReverberationQSO}).
The time-delays between the continuum flux of AGN and ultra-luminous X-ray sources at different wavelengths (e.g., \citealt{Kara+2016_AGN_Xray_TimeLags}, \citealt{Kara+2020_ULX_Xray_TimeElags})
presumably provide clues as to the internal structure of accretion disks.
Yet another example is the identification of lensed quasars and lensed supernovae (e.g.,
\citealt{Young+1980_Q0957+561_LensedQuasar};
\citealt{Goobar+2017_PTF16geu_LensedIaSN}) and the measurement of the time delay between their multiple images. Time-delay measurements in these systems may help in recovering the mass of the lensing galaxies (e.g., \citealt{Maoz+Rix1993_LensedQSO_Stat}, \citealt{Treu+2006_LensedGalSurveyIII_GalProfile});
studying the host galaxies of quasars (e.g., \citealt{Sharon+2012_SourcePlaneReconstructionLensedGalaxy});
probing the galaxy mass evolution (e.g., \citealt{Ofek+2003_LensedQSOz_GalEvolution});
measuring cosmological parameters (e.g., \citealt{Kochanek1992_LensedQSO_Lambda}, \citealt{Maoz+Rix1993_LensedQSO_Stat});
studying the accretion disk around massive black holes (e.g., \citealt{Kochanek2004_QSO_AccDisk_Microlensing}, \citealt{Morgan+2010_QSO_AccreationDiskSize_MassBH},
\citealt{Chan+2020_QSO_AccreationDiskSize_QSO_Microlensing});
and measuring the Hubble constant (e.g.,
\citealt{Refsdal1964_HubbleConstant_FromLensingTimeDelay},
\citealt{Saha+2006_HubbleTime_LensedQSO_TimeDelay}, \citealt{Oguri2007_LensTimeDelaySample_H0}, \citealt{Birrer+Treu2020_TimeDelay_H0_Strategies}, \citealt{Wong+2020_LensedQuasar_H0_measurment}),
although the latter is limited by the knowledge of the mean convergence at the Einstein radius (e.g., \citealt{Kochanek2020_LensedQSO_TimeDElay_MeasuringH0_OverConstr}, \citealt{Blum+2020_LensedQuasars_H0_Core}).
Another possible case in which we expect for a time delay between different parts of the emitted radiation is in the millimeter-wavelength emission from the silhouette around massive black holes (\citealt{Hadar+2020_BlackHole_PhotonRingAutoCorrelation}).

These targets share several complicating factors.
First, in many cases, the time-delayed components are unresolved
(either spatially or in wavelength-space).
For example, in lensed quasars, the typical image separation
is of the order of $0.5''$, just below the seeing disk induced by the Earth's atmosphere.
This fact makes it difficult to detect and measure the time delays of lensed quasars with small angular separations. For lensed Type Ia supernovae (SNe), the situation may be somewhat easier, as the light curve of the family of Type Ia SNe light curves is known and can be modeled using a small number of free parameters (e.g., \citealt{Dhawan+2019_TimeDelay_Lensed_SNIa_PTF}; \citealt{Bag+2020_MethodTimeDelay_LensedSN}).

Similarly, in order to measure reverberation time delays it is necessary to spectroscopically resolve the delayed components (i.e., the broad line region from the continuum).
Both high-resolution and spectroscopic observations are considerably more expensive compared with seeing dominated broad-band synoptic observations.

A second complication is that quasar light curves are generated by a red-noise process -- i.e.,
the power spectrum of quasar light curves is well described by a steep power-law
(e.g., \citealt{Mushotzky+2011_AGN_PowerSpectra_KeplerLC},
\citealt{Edelson+2014_Kepler_AGN_Zw229-15_5day_timescale},
\citealt{Kasliwal+2015_IsAGNLCConsistentWithDampedRandomWalk_AGN_PowerSpectrum},
\citealt{Smith+2018_AGN_KeplerLC_PowerSpectrum}).
This means that quasar light curves are highly correlated in the time domain, and their variability amplitude decreases with frequency.
Ignoring this fact can bias time-delay measurements 
and cause severe underestimation of the time-delay uncertainties (e.g., \citealt{Ofek+2003_HE1104-1805_TimeDelay}).
This red-noise process can also generate peaks in the power spectrum that are sometimes mistakenly identified as a periodicity (as demonstrated in \citealt{Cenko+2010_GRB090709A_LackOfPeriodicity}).

\cite{Geiger+Schneider1996_LensedQuasarTimeDelay_AutoCorrelationFunction} investigated the possibility of measuring
the time delay of a lensed quasar from its combined light curve using the auto-correlation function.
They concluded that this is a viable method in cases where the flux ratio is close to unity.
They also suggested that some of the problems of this method can be overcome by adding a few observations of the flux ratio (i.e., resolved observations of the system).
\cite{Pindor2005_LensedQuasarTimeDElay_AutoCorrelationFunction}
suggested the use of dispersion statistics
to recover the time delay of a combined light curve of a lensed quasar.
As discussed by \cite{Pindor2005_LensedQuasarTimeDElay_AutoCorrelationFunction} and explained here, this is not always possible and requires a statistical model for the quasar variability.
Recently, \cite{Shu+2020_LensedQuasarTimeDElay_AutoCorrelationFunction} demonstrated that a modified version
of the  auto-correlation function can recover the time delays
of real systems with a success rate of about 20\%.

Here, and in a companion paper (Springer \& Ofek; hereafter Paper II), we derive methodologies relevant for time-delay measurements.
In this paper, we discuss the scenario of having
observations of the combined flux of several time-shifted light curves.
This is relevant for time-delayed sources that are either spatially or spectrally unresolved.
We show that, given that quasar light-curves are generated from a red-noise process (or any known process with power over a broad band of frequencies), it is possible
to measure the time delay from the combined flux light curve.
In Paper II, we describe a method for the identification of lensed quasars and for their time-delay measurement.
The method described in Paper II uses, in addition to the total-flux measurements, a measurement of the astrometric
position of the center of light of all of the combined images.
As the second method uses additional information, it is more powerful than the method
described in this paper. However, the second method is not suitable for quasar reverberation mapping.

Our approach is different than the one outlined by
\cite{Geiger+Schneider1996_LensedQuasarTimeDelay_AutoCorrelationFunction}, \cite{Pindor2005_LensedQuasarTimeDElay_AutoCorrelationFunction}, and \cite{Shu+2020_LensedQuasarTimeDElay_AutoCorrelationFunction}
in several respects.
First, we use the fact that we have a statistical model
for the light curve of quasars. Ignoring the fact that quasar light curves are highly correlated in time will lead
to underestimation of the uncertainties.
Second, we rigorously derive the relevant likelihood function,
and 
third, the likelihood function we derive is different from the
auto-correlation function and the $\chi^{2}$ statistic.

The structure of this paper is as follows.
In \S\ref{sec:description}, we describe the problem we would like to solve and provide a qualitative intuition for why a solution may exist.
In \S\ref{sec:method} we derive the likelihood function required in order to apply our method.
In \S\ref{sec:practical}, we discuss practical considerations,
while in \S\ref{sec:simulations} we test the method
on simulated data.
In \S\ref{sec:code}, we describe our code, and we conclude 
in \S\ref{sec:disc}.

\section{Schematic description of the method}
\label{sec:description}

We are interested in studying a light curve
that is possibly composed of two or more identical copies that are time
shifted with respect to each other, with each of the copies multiplied by a different
flux normalization factor.
For simplicity, here we discuss the case of two copies, while the general case of multiple copies is derived in Appendix~\ref{App:multi_image}.
We will call the identical copies: the
{\it original light curve} or
{\it source light curve},
and the {\it time-delayed source light curve},
while the linear combination of the two will be called the {\it combined light curve}.
Given that we have access only to the combined light curve, we are interested
in addressing the following questions:
Is it possible to test if a combined light curve was generated
by such a process, or from an alternative process
in which we observed only the source light curve
(i.e., no time delay)? If so, is it possible to estimate the time delay and flux ratio between the copies?

In the most general case, the answer to these questions is {\it no}.
To grasp this, we can consider the case of a source light curve that is a periodic function. A linear combination of a periodic function and its time-shifted version may be explained using an infinite number of
time delays.
It is clear that a statistical model of the source light curve is needed to be able to differentiate between the two hypotheses mentioned above. Furthermore, a successful hypothesis testing may be possible only for cases in which there is power over a wide range of frequencies.

Quasar light curves are generated by a roughly known statistical process that has power in a large number of frequencies. Currently, the best statistical description
of quasar light curves is that in Fourier space, they have a power-law, or broken power-law,
red\footnote{By red power-spectrum we mean that lower frequencies have higher amplitudes.}
power-spectrum of the form $\omega^{-\gamma}$, where $\omega$ is the angular frequency
and $\gamma$ is the power-law index.
Previous analysis of quasar light curves suggest $\gamma$ values in the range of $1.5$ to $3.5$
(e.g., \citealt{Mushotzky+2011_AGN_PowerSpectra_KeplerLC}; \citealt{Zu+2013_IsQuasarVariability_DampedRandomWalk_RedPowerSpectrum}; \citealt{Kasliwal+2015_IsAGNLCConsistentWithDampedRandomWalk_AGN_PowerSpectrum}; \citealt{Smith+2018_AGN_KeplerLC_PowerSpectrum}).
Given the difficulty of measuring $\gamma$ (see also \S\ref{sec:PLfit}), the measurements typically have large uncertainties,
and it is not clear if all quasar light curves are generated from
a universal distribution (i.e., having the same power-law index $\gamma$).
For simplicity, here we test our method on
a single power-law model, but an extension
to other models is straightforward.

Given these facts, there is hope that it is possible to
derive a method that will allow us to test if a light curve
was generated from a linear combination of a source light curve with a time-delayed
source light curve and to simultaneously measure this time delay.
At the heart of our approach is the use of the knowledge of the rough probability distribution of the source light curves, from which we may then derive a statistical model for the distribution of the combined light curves.

\section{Method formal derivation}
\label{sec:method}

We assume that the original light curve is generated by a
process for which we have a statistical model for its power-spectrum.
Given this assumption, we would like to test
the alternative hypothesis ($H_{1}$) that the observed light curve is composed of two time-shifted copies of an original light curve, against the null hypothesis ($H_{0}$) that the observed light curve is composed of a single copy of the original light curve. Next, if the alternative hypothesis is true, we would like to estimate the time-delay between the copies.
The only observable here is the total flux of the combined light curve.

In order to achieve these goals, we are interested in expressions for the likelihood
of the observations given the model and the free parameters,
where the free parameters may include the
time delay between the images, the flux ratio between the images,
and the properties of the source power spectrum (e.g., $\gamma$).
This likelihood can be used to measure the free parameters,
or to decide if the combined light curve is composed of time-shifted copies.
In the latter case, the decision is made against the null hypothesis
(e.g., that the quasar is not lensed), using the likelihood ratio test
(\citealt{Neyman+Pearson1933_HypothesisTesing}).

In the following subsections, we present the derivation for this likelihood.
In \S\ref{sec:model} we present the statistical model for the flux.
In \S\ref{sec:FTvsTD} we discuss the two approaches to the problem - i.e., working in the Fourier space or time domain.
In \S\ref{sec:stat_flux}, we derive the 
likelihood function of the observations, in Fourier space,
given the unknown parameters,
while in \S\ref{sec:like_TD} we do this in the time domain
with the full covariance matrix.
In \S\ref{sec:kernel}, we extend this to the case in which the time-delay is described by a response kernel
(i.e., convolution with a finite-width function rather than a delta function).
In \S\ref{sec:PLfit}, we comment on a related question of how to fit
a power-law model to a power spectrum.

\subsection{Flux formation model}
\label{sec:model}

For simplicity, we present the derivation of our method for a light curve that is composed of two time-shifted
light curves.
In Appendix~\ref{App:multi_image},
we extend this to the general case of multiple time-shifted light curves.

For the two-image case, the model for the total observed flux is given by:
\begin{eqnarray}
    F(t) &=& \phi(t) + \epsilon_{F}(t) \nonumber \\
         &=&\alpha_0 + \alpha_1 f(t) + \alpha_2 f(t+\tau) + \epsilon_{F}(t),
    \label{eq:F}
\end{eqnarray}
where $\phi(t)$ is the original total flux (without observational noise), $\alpha_0$ is the flux of a non-variable component (e.g., lensing galaxy),
$f(t)$ is the source light curve
as a function of the time $t$, $\tau$ is the time delay between image 1 and 2,
$\alpha_{i}$ is proportional to the mean flux of the $i$-th image,
and $\epsilon_{F}(t)$ is the noise in the total flux
measurement. We assume that the observations are background-noise dominated
and that  $\epsilon_{F}(t)$ is an independent and 
identically distributed (i.i.d.) random Gaussian vector with a
per-component variance $\sigma_{F}^2$. $\alpha_{2}/\alpha_{1}$ can be regarded as the mean flux ratio between the two images/components.
The requirement for $\epsilon_F$ to be i.i.d. can be relaxed, however, this will introduce the complication of requiring the full covariance matrix (using the formalism discussed in \S\ref{sec:like_TD}).

We note that, in practice, even if we set the measurement error to zero, the light curves of the various images (properly normalized by $\alpha_i$) may not be identical.
A leading problem, in the case of lensed quasars, is variability due to microlensing by individual stars in a lensing galaxy (e.g., \citealt{Wambsganss+2000_LensedQuasar_Q0957+561_Microlensing}; \citealt{Wambsganss2001_LensedQuasars_Microlensing};
\citealt{Ofek+2003_HE1104-1805_TimeDelay}; \citealt{Goldstein+2018_LensedSupernovaeMicrolensingImpact}).
To simplify the
analysis, we absorb all these variations into the Gaussian noise term $\epsilon_{F}$ (see also \S\ref{sec:FluxErrors}).

The particular light curve of the observed quasar's first image $f(t)$ is unknown {\it a priori}. We, therefore, model it statistically. Following the definition of the Fourier transform
\begin{equation}
\mathcal{F}[f(t)]\equiv\widehat{f}(w) = \int_{-\infty}^\infty f(t)e^{i\omega t} dt,
\end{equation}
where we denote the Fourier Transform by $\mathcal{F}$ or the hat sign above the function.
We assume that in the frequency domain, quasar light curves have the following Gaussian (Normal; $N$) distribution:
\begin{equation}
    \widehat{f}(\omega) \sim N(0, \sigma_{\widehat{f}}^2(\omega)),
    \label{eq:f_FT_normal}
\end{equation}
where at each frequency $\omega$, $\widehat{f}(\omega)$ is a complex number with independent real and imaginary parts, each having a zero mean and a variance of $\sigma_{\widehat{f}}^2(\omega)/2$.
We are using the tilde sign ($\sim$) here to denote the distribution of a random variable.
We assume here that $\widehat{f}(\omega)$ is also statistically independent at different frequencies, has a zero mean, $\operatorname{E}\left[\widehat{f}(\omega)\right] = 0$ for frequencies $|\omega| > 0$, and that the variance function has the following power-law shape for $\omega\ne0$:
\begin{equation}
    \sigma_{\widehat{f}}^2(\omega) = \operatorname{E}\left[\widehat{f}(\omega)\overline{\widehat{f}}(\omega)\right] =
    \operatorname{E}\left[| \widehat{f}(\omega) |^{2}\right] = 
    |\omega|^{-\gamma}.
    \label{eq:PowerLawModel}
\end{equation}
Here the bar sign above the hat symbol indicates a complex conjugation applied after the Fourier transform.
We note that the power spectrum in this equation has no normalization -- i.e., the variance at unit frequency is one.
We find that it is a bit more convenient to let the amplitude normalization of the power spectrum enter
through the flux normalization of the target.
Note that we use $\sigma_{\widehat{f}}^2$ to denote the
power spectrum of $f$, while we use the notation $\widehat{\sigma}_{f}$ to denote the Fourier Transform
of the errors in $f$.

This assumption regarding the power-law power-spectrum of the light curves of quasars is supported by observations (see \S\ref{sec:description}; e.g., \citealt{Markowitz+2003_Quasars_AGN_Xray_powerspectra}; \citealt{Mushotzky+2011_AGN_PowerSpectra_KeplerLC}; \citealt{Smith+2018_AGN_KeplerLC_PowerSpectrum}), which currently suggest that $\gamma\approx1.5$--$3.5$.
Given that the accurate characterization of the underlying process from which quasar
light curves are generated is not fully known, the simple power-law representation provides a good approximation.
In \S\ref{sec:sensitivity}, we test the sensitivity of our solution
to this assumption.
Furthermore, it is straightforward to use different models instead of the one assumed in Equation~\ref{eq:PowerLawModel}
(e.g., a broken power-law or an SN light curve).

\subsection{Fourier space vs. time domain}
\label{sec:FTvsTD}

An important practical consideration that we now discuss is:
should we work in Fourier space or in the time domain?
At first glance it is tempting to
work in Fourier space.
The main reason is that, in the quasar
light curve model (Eq.~\ref{eq:PowerLawModel})
different frequencies are independent
and therefore the covariance matrix is diagonal.
However, for many practical reasons this is not the case,
and there are several problems:
(i) For unevenly spaced data the different frequencies become correlated;
(ii) even in the case of evenly spaced data, the observations do not cover
all the frequencies in which the model has power,
and the window function introduces correlations between different frequencies;
(iii) applying the Fast Fourier Transform assumes that our
time series has cyclic boundary conditions which is clearly not the case.

By working in the time domain, and taking into
account the full covariance matrix, we can overcome these problems.
The price we will have to pay for calculating
the full covariance matrix is that we will have to
use matrix inversion (which is relatively slow),
and that we will have to deal with diverging integrals
when we calculate the covariance matrix (see \S\ref{sec:FitFullCov}).

Here we present the solution to the problem
both in Fourier space with a diagonal covariance matrix (approximate solution)
and in the time domain with the full covariance matrix (accurate solution).

We note that our Fourier space solution will not work well
without some heuristic (but justified) massaging to the data.
The main method that can partially fix some of the problems
outlined above, is the popular end-matching technique
(e.g., \citealt{Uttley+2002_PowerSpectraAGN_RXTE}).
This includes removing a linear trend from the light curve
such that the first and last point will have the same flux
(i.e., making the time series appear more cyclic).

\subsection{The likelihood of a light curve given a model in Fourier space}
\label{sec:stat_flux}

Given the flux formation model, here, we derive
the probability of the flux given the unknown parameters (e.g., $\tau$, $\alpha_{i}$, $\gamma$)
in Fourier space.
We start by expressing the noiseless total flux (Equation~\ref{eq:F}) in the frequency domain,
\begin{eqnarray}
    \widehat{\phi}(\omega) &\equiv& \mathcal{F}\left[\alpha_0 + \alpha_1 f(t) + \alpha_2f(t+\tau)\right]  \nonumber \\
    &=& \alpha_0\delta(\omega) + (\alpha_1+\alpha_2 e^{i\omega\tau})\widehat{f}(\omega).
    \label{eq:Phi_of_omega}
\end{eqnarray}
Here $\delta(\omega)$ is the standard Dirac delta function with a definite integral of $1$ over all intervals, including $\omega = 0$, and it is equal zero when $\omega \neq 0$.
In the following, expressions in the frequency domain will be written such that they are valid for $|\omega| > 0$ unless stated otherwise. Therefore, in this case, we drop the temporally constant term $\alpha_0\delta(\omega)$.
This is valid, as the power at zero frequency is analogous to a constant, which we may subtract from the entire light curve. The resulting mean of the noiseless total flux is
\begin{equation}
    \operatorname{E}\left[\widehat{\phi}(\omega)\right] = 0,
\label{eq:Exp_phi}
\end{equation}
while the expectation value of the power spectrum of the light curve is
\begin{eqnarray}    
    \label{eq:Sigma_phi}
     \operatorname{E}\left[\widehat{\phi}(\omega)\overline{\widehat{\phi}}(\omega)\right] &\equiv& \Sigma_{\phi}(\omega) \\ \nonumber
     &=& \frac{\alpha_1^2+\alpha_2^2+2\alpha_1\alpha_2 \cos(\omega\tau)}{|\omega|^\gamma}.
\end{eqnarray}
Note that we use the small $\Sigma$ symbol here and in the rest of the text to denote covariance functions and covariance matrices,
while $\Sigma(\omega)$ is regarded as the diagonal of the covariance matrix.

The noisy total observed flux, in Fourier space,
is given by
\begin{equation}
    \widehat{F}(\omega) = (\alpha_1+\alpha_2 e^{i\omega\tau})\widehat{f}(\omega) + \widehat{\epsilon}_F(\omega).
    \label{eq:F_of_omega}
\end{equation}
From Equations~\ref{eq:f_FT_normal} and \ref{eq:Exp_phi}, we get that
$\operatorname{E}\left[\widehat{F}(\omega)\right] = 0$, while the variance is
\begin{equation}
    \operatorname{E}\left[\widehat{F}(\omega)\overline{\widehat{F}}(\omega)\right] \equiv \Sigma_{F}(\omega) = \Sigma_{\phi}(\omega) + \widehat{\sigma}_F^2.
    \label{eq:F_var}
\end{equation}
Equation~\ref{eq:F_var} has several important properties. First, it is symmetric around $\tau=0$
with respect to $\tau$, and it is also symmetric with respect to exchanging the role
of $\alpha_{1}$ and $\alpha_{2}$.
Given a particular observation $F(t)$ and its frequency representation $\widehat{F}(\omega)$, the log-probability of observing $F(t)$ given the model parameters is
\begin{eqnarray}
\label{eq:log_P_flux}
    \ln P(\widehat{F} | \tau, \alpha_i, \gamma) = -\frac{1}{2}\ln \det[2\pi\Sigma_{F}] - 
    \sum_{\omega}\frac{|\widehat{F}(\omega)|^2}{2\Sigma_F(\omega)}.
    \label{eq:logP_F_par}
\end{eqnarray}
Equation~\ref{eq:logP_F_par} results from the expression for the log-probability of a multivariate normal distribution (see appendix \ref{App:cmvn}).
Here we treat $\widehat{F}(\omega)$ as a vector sampled at particular frequencies $\omega$
(i.e., $|\widehat{F}(\omega)|^{2}$ is the observed power spectrum) and $\Sigma_{F}(\omega)$ as a diagonal matrix.
The det operator is the determinant.
In practice, in order to avoid multiplications of small numbers,
we should use summation rather than the product:
\begin{equation}
\ln(\det[2\pi\Sigma_{F}]) = \sum_{\omega}{\ln{[2\pi\Sigma_{F}(\omega)]}}.
\end{equation}

We note that in the general case, the distribution in frequency space may contain correlations between frequencies (see \S\ref{sec:like_TD}). In such a case, the appropriate covariance matrix $\Sigma_{F}$ would contain non-diagonal terms and the general expression for a non-diagonal multi-variate normal distribution should be used (see Appendix~\ref{App:cmvn} and \S\ref{sec:FitFullCov}).
When $\Sigma_{F}$ is a matrix, the $\ln{\rm det}$ should be calculated using the prescription in Appendix~\ref{sec:logdet}.
The formula analogous to Equation \ref{eq:logP_F_par} that is suitable for an arbitrary number of images is given in Appendix~\ref{App:multi_image}.

In Figure \ref{fig:power}, we plot
the expectation value of a combined light curve power spectrum (i.e., $\Sigma_{F}(\omega)$; Equation~\ref{eq:F_var})
for particular choices of the parameters.
We see qualitatively that the model predicts that fluctuations in the combined flux $F(t)$ occurring on time scales that correspond to angular frequencies $\omega \approx 2\pi n/\tau$ (for integer $n\geq 0$) are enhanced in the observed total flux $F(t)$, while those occurring at $\omega \approx (2\pi n+\pi)/\tau$ are suppressed.
\begin{figure}
\centerline{\includegraphics[width=8cm]{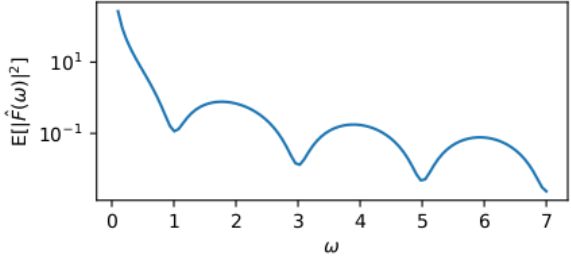}}
\caption{Expected total observed power (i.e., Eqs.~\ref{eq:Sigma_phi} and \ref{eq:F_var}) for model parameters $\tau=\pi$, $\alpha_1=1$, $\alpha_2=2/3$, $\gamma=2$ and $\widehat{\sigma}_F=1/10$.
\label{fig:power}}
\end{figure}

\subsection{The likelihood function in the time domain with the full covariance}
\label{sec:like_TD}

Following the derivation in \S\ref{sec:stat_flux},
and the expression for the log-probability of a multivariate normal distribution (see Appendix \ref{App:cmvn})
we can write the log-likelihood in the time domain:
\begin{eqnarray}
    \ln P(F | \tau, \alpha_i, \gamma) 
    = -\frac{1}{2}\ln \det[2\pi\Sigma_{T}] \notag \\ 
      -\frac{1}{2}(F(t)-\mu_{F})^{T} \Sigma_{T}^{-1} (F(t)-\mu_{F}).
    \label{eq:LL_TD}
\end{eqnarray}
Here upper-script $T$ denote the transpose operator, $\mu_{F}$ is the expectancy value of $F(t)$ (approximated using the mean of $\mu_{F}$), and $\Sigma_{T}$ is the covariance matrix between all the pairs in $F(t)$ in the time domain.
In order to avoid the multiplication of small numbers, practical calculation of the log of the determinant is described in Appendix~\ref{sec:logdet}.

The elements of $\Sigma_{T}$
between two times $t_{j}$ and $t_{k}$
are:
\begin{equation}
    \Sigma_{T}(t_{j},t_{k}) =
    \operatorname{E}[(F(t_{j})-\mu_{F}){(F(t_{k})-\mu_{F})}].
\end{equation}
The Wiener–Khinchin theorem relates the auto-correlation function
to the power spectral density via the Fourier transform.
Therefore,
\begin{eqnarray}
    \Sigma_{T}(t_{j},t_{k})
    &=& \int_{-\infty}^{+\infty}{\Sigma_{F}e^{-i\omega(t_{j}-t_{k})} d\omega} \notag\\
    &=& 2\int_{0}^{\infty}{\Sigma_{F} \cos{(\omega[t_{j}-t_{k}])} d\omega}\notag \\
    &=& 2\int_{0}^{\infty}{ \frac{\alpha_{1}^{2} + \alpha_{2}^{2} + 2\alpha_{1}\alpha_{2} \cos{(\omega\tau)} }{\omega^{\gamma}} } \notag\\
    & & \times\cos{(\omega[t_{j}-t_{k}])} d\omega + \delta(t_{j}-t_{k})\sigma_{F}^{2},
    \label{eq:SigmaT_t}
\end{eqnarray}
where the second line is valid since 
$\Sigma_{F}$ is a real valued function (and hence the auto-correlation is symmetric around 0).
The addition of the last term
indicates that the observational variance of the flux should be added to the diagonal of the covariance matrix.

An apparent problem is that this integral is diverging for any value of $\gamma$.
In Appendix~\ref{sec:AutoCorr},
we present two methods to deal with this problem.
The first is to start the integration at a small positive $\omega_{0}$,
while the second is to re-write this integral in terms of converging integrals (for $3>\gamma>1$) and one diverging integral
that depends only on $\gamma$.
Removing this diverging integral (Equation~\ref{eq:G0}) has no major effect on the likelihood (compared to the likelihood of the null hypothesis).
These procedures are verified on simulations, and its results are presented in \S\ref{sec:simulations}.
In Appendix~\ref{sec:AutoCorr}, we also show that the two approaches work well.

We note that there is a physical
reason to why it is justified to ignore the diverging integral,
or to start the integration at a small $\omega_{0}$.
The reason is that so far we assumed
a power-law power spectrum that
continues to zero frequency.
This is nonphysical, as we expect that
the power-law power-spectrum will have a cutoff at some low frequency.
This cutoff frequency is in most cases unknown.
This means that $\Sigma_{T}$ is known
up to a constant.
The nature of this unknown constant
is analogous to the true DC of the light curve
and some very low frequency variations (e.g., a trend).

\subsection{Time delay with a response kernel}
\label{sec:kernel}

For some applications, we may be interested in a slightly different model,
in which the delayed signal is smeared with some temporal response kernel $K(t)$.
An obvious application is measuring the reverberation time delay of quasars.
In this case, the statistical model in Equation~\ref{eq:Phi_of_omega}
will become
\begin{eqnarray}
    \widehat{\phi}(\omega) &\equiv& \mathcal{F}\left[\alpha_0 + \alpha_1 f(t) + \alpha_2f(t+\tau)* K(t)\right]  \nonumber \\
    &=& \alpha_0\delta(\omega) + (\alpha_1+\alpha_2 \widehat{K} e^{i\omega\tau})\widehat{f}(\omega).
    \label{eq:Phi_of_omega_K}
\end{eqnarray}
Here, $*$ is the convolution operator,
and the expectation value of the power spectrum becomes
\begin{align}    
     \operatorname{E}\left[|\widehat{\phi}(\omega)|^{2}\right] \equiv \Sigma_{\phi}(\omega) & \\ 
     =\frac{\alpha_1^2+\alpha_2^2 |\widehat{K}|^{2} +2\alpha_1\alpha_2[ \cos(\omega\tau) {\rm Re}(\widehat{K}) -  \sin(\omega\tau) {\rm Im}(\widehat{K})] }{|\omega|^\gamma}&.
    \label{eq:phi_var_K}
\end{align}
For $K(t)=\delta(t)$, Equation~\ref{eq:phi_var_K} reduces to Equation~\ref{eq:Sigma_phi}.

In practice, Equation~\ref{eq:phi_var_K} involves a large number of free parameters
(i.e., the elements of $K$).
In principle, one can reduce the number of free parameters by assuming $K$ has a specific shape
(e.g., a Gaussian).

\subsection{A comment about power-law fitting}
\label{sec:PLfit}

We note that naively one may be tempted to
fit Equation~\ref{eq:Sigma_phi} to the data
using a $\chi^{2}$ minimization
of the form
\begin{equation}
    \chi^{2}=\sum_{\omega}{\frac{(|\widehat{F}(\omega)|^{2} - \Sigma_{F} )^{2}}{\widehat{\sigma}_{F}^{2}}}.
    \label{eq:naive_chi2}
\end{equation}
Here $\Sigma_{F}$ represents any model (e.g., a power-law or a broken power-law).
In fact, such an approach is commonly used in the literature when attempting to fit
a power-law to a power spectrum (e.g., \citealt{Fougere1985_SpectrumRedNoiseAnalysis_Accuracy};
\citealt{Uttley+2002_PowerSpectraAGN_RXTE}; \citealt{Smith+2018_AGN_KeplerLC_PowerSpectrum}).

However, as discussed in \S\ref{sec:stat_flux}, the expectation value
of the Fourier transform of the process is zero, while its variance is given by $\Sigma_{F}$.
Therefore, using Equation~\ref{eq:naive_chi2} will result in effectively using
the wrong probability distribution for the data.
While Equation~\ref{eq:naive_chi2} assumes that the measurements have
Gaussian noise in the power spectrum, we assume that the data
has Gaussian noise in Fourier space.

Another difference between Equation~\ref{eq:naive_chi2} and our approach is that there
is an additional additive term in the likelihood. This term depends on the model parameters
and should therefore not be ignored in a fitting process.
To summarize, using Equation~\ref{eq:logP_F_par} when fitting a model (e.g., a power-law)
to the power spectrum will result in more accurate results,
compared to using Equation~\ref{eq:naive_chi2}.

\section{Practical considerations}
\label{sec:practical}

Here we discuss several practical issues, including
aliasing and red-noise leak (\S\ref{sec:RedLeak}),
unevenly spaced data (\S\ref{sec:unevenlyspaced}),
flux uncertainties (\S\ref{sec:FluxErrors}),
and multi-band data (\S\ref{sec:MultiBand}).
Finally, in \S\ref{sec:broadband}, we discuss
the expected $\alpha_2/\alpha_1$ for quasar reverberation mapping
in some realistic scenarios.

\subsection{Red-noise leak and aliasing}
\label{sec:RedLeak}

When working in Fourier space without the full covariance, there are several phenomena, including red-noise leak and aliasing, that may influence the results
(e.g., 
\citealt{Deeter+Boynton1982_EstimatingRedPowerSpectraI};
\citealt{Edelson+Nandra1999_XrayPowerSpectrum_AGN_NGC3516};
\citealt{Uttley+2002_PowerSpectraAGN_RXTE};
\citealt{Press+2002_Book_NumericalRecepies}).
We can overcome some of these problems by using the full covariance
(e.g., \S\ref{sec:like_TD}).

The problem of aliasing is related to the transfer of
power from frequencies $f_{\rm Nyq}+\Delta{f}$ to $f_{\rm Nyq}-\Delta{f}$,
where $f_{\rm Nyq}$, the \textit{Nyquist frequency}, is half the sampling frequency.
While our model assumes that the data has
a cutoff above the sampling frequency, in reality,
quasars may have power at higher frequencies.
To deal with this,
in our simulations, we generated the light curves from power spectra that were extended to frequencies which are ten times higher than the desired (observed) sampling frequency.
Since in the case of quasars, we are mainly dealing with
red-power-law power spectra,
there is not much power above the sampling frequency,
and this technique is a reasonable approximation.
The red-noise leakage effect happens when low frequencies (below the lowest frequency in the data) leak into higher frequencies.

A simple way to partially deal with these problems
is to use the end-matching technique
(e.g., \citealt{Fougere1985_SpectrumRedNoiseAnalysis_Accuracy},
\citealt{Uttley+2002_PowerSpectraAGN_RXTE}).
The idea is to remove a slope from the light curve
such that the start point and end point of the light curve will have the same flux.
Indeed, when we apply the end-matching process
to our simulations,
the Fourier-space results improve
(see \S\ref{sec:simulations}).

\subsection{Unevenly spaced data}
\label{sec:unevenlyspaced}

In many cases, astronomical time series are unevenly spaced.
Spectral analysis of unevenly spaced data is complicated
by the fact that the power spectrum
of an unevenly spaced time series is equal to
the power spectrum of the continuous time series
convolved with the power spectrum of the window function
(e.g., \citealt{Deeming1975_periodogram}).

Since quasars have a red power spectrum their variability
amplitude on short time scales is small.
Therefore, interpolations of an unevenly spaced time series
into an evenly spaced time series seems reasonable.
However, such interpolation
is still effected by aliasing.
Our full covariance method (\S\ref{sec:like_TD})
removes the need for using interpolation.
In this case we need to calculate the covariance matrix $\Sigma_{T}(t_{i},t_{j})$
for all possible pairs of observations.
Indeed, our simulations (\S\ref{sec:simulations}) suggests that the full covariance method provides better results compared with the Fourier domain method.

We note that some past, present, and future space missions (e.g., {\it Kepler}, \citealt{Borucki+2010LeplerMission_IntroFirstResults}; {\it TESS}, \citealt{Ricker+2015_TESS_MissionInstrument};
and {\it ULTRASAT}, \citealt{Sagiv+2014_ULTRASAT})
provide continuous monitoring of the sources
and hence (almost) evenly spaced time series.
The analysis of evenly spaced data sets is simplified,
compared to unevenly spaced datasets, due to the fact that $\Sigma_{T}$ is easier to calculate.
Specifically, for unevenly space data, in order to calculate $\Sigma_{T}$ we have to evaluate the integral in Equation~\ref{eq:SigmaT_t} $\approx N^{2}/2$ times,
while for evenly spaced data we
can evaluate this integral only $\approx N$ times.
This is because, in the evenly spaced case, each diagonal in $\Sigma_{T}$ contains the same integral (i.e., a Toeplitz matrix).

\subsection{Model and flux uncertainties}
\label{sec:FluxErrors}

The typical expected photometric precision in some synoptic sky surveys
is better than $0.02$\,mag (e.g., \citealt{Padmanabhan+2008_SDSS_ImprovedPhotometricCalibration};
 \citealt{Ofek+2012_PTF_photCat}; \citealt{Schlafly+2012_PS1_PhotometricCalibration};
 \citealt{Masci+2019_ZTF_Pipeline}).
However, in the case of lensed quasars, additional sources of noise
like microlensing maybe relevant (e.g., \citealt{Wambsganss+2000_LensedQuasar_Q0957+561_Microlensing}; \citealt{Eigenbrod+2008_LensedQuasar_Q2237+0305_Microlensing}).
Furthermore, this noise may be correlated and have some specific time
scales corresponding to caustic crossings ($\sim3$\,months) and Einstein radius crossing ($\sim10$\,years; e.g., \citealt{Wambsganss2001_LensedQuasars_Microlensing}).
\cite{Ofek+2003_HE1104-1805_TimeDelay} found
residual variability in the HE1104$-$1805 lensed quasar light curve,
with peak-to-peak variations of up to 30\%,
and with rms of about 0.07\,mag.
Such additional variations need to be taken into account.
The simplest approximation is to absorb these additional noises into $\sigma_{F}$.

\subsection{Multi-band data}
\label{sec:MultiBand}

Some sky surveys produce multi-band light curves.
Using information from multiple bands is important, as it may help
us reduce the gaps between the observations and increase the number of photometric data points.
It is possible to measure the time delay using all the bands simultaneously.
In this case, we need to maximize the global likelihood of
all the bands, so that we can assume that all the bands share the same time delay
but have different $\alpha_{1}$ and $\alpha_{2}$ values.

\subsection{$\alpha_2/\alpha_1$ in broad-band reverberation mapping}
\label{sec:broadband}

As our simulations suggest,
our method can work well when $\alpha_{2}/\alpha_{1} \gtorder 0.1$ (see \S\ref{sec:simulations}).
For quasar reverberation mapping, $\alpha_{2}/\alpha_{1}$ is
approximately the equivalent width of the emission lines
in the observed band
divided by the band width.
We use the spectral fitting tools in \cite{Ofek2014_MAAT} to measure the equivalent width
of spectral lines in the Sloan Digital Sky Survey (\citealt{York+2000_SDSS_Summary}) quasar composite spectra of \cite{VandenBerk2001_CompositeQSO_Spec_SDSS}.
We find the equivalent width of some prominent lines to be:
$\approx$150\AA~(H$\alpha$; 6564\AA),
30\AA~(H$\beta$; 4862\AA),
20\AA~(\ion{Mg}{2}; 2799\AA),
20\AA~(\ion{C}{3}; 1909),
25\AA~(\ion{C}{4}; 1550\AA),
9\AA~(\ion{O}{4}$+$\ion{Si}{4}; 1400\AA),
90\AA~(Ly$\alpha$; 1216\AA).

For a {\it Kepler}-like transmission band ($\approx4000$\AA~width) and low redshift quasar,
the H$\alpha$ and H$\beta$ lines are in the band and will result in an $\alpha_{2}/\alpha_{1}\approx0.05$.
This is low, and likely hard to detect using our method.
For a quasar at $z\approx3$, the Ly$\alpha$ line
will be in the $g$-band and will result in an $\alpha_2/\alpha_1\approx0.3$.
Similarly, for a quasar at $z\approx1$, the Ly$\alpha$ line will be in
the {\it ULTRASAT} NUV band (\citealt{Sagiv+2014_ULTRASAT}),
and will result in an $\alpha_2/\alpha_1\approx0.3$.

\section{Simulations}
\label{sec:simulations}

In order to simulate a red-power-spectrum time series, we use the prescription
described in Appendix~\ref{app:cycsim}.
We use this method to generate light curves
with non-cyclic boundary conditions.
Another important detail is that in order to generate
realistic simulations we need
to make sure that the simulated light curves have
power at frequencies below and above the
lowest and highest observed frequencies (see \S\ref{sec:RedLeak}).

We expect that the method that uses the full covariance in the time domain, will be superior to the Fourier space method.
One disadvantage of our time domain method
is that it requires matrix inversion and some integral evaluations, and
therefore it is about 100 times slower to calculate than
the Fourier domain method.
We recommend using the full-covariance method
for any real application.

Here, whenever possible,
we use the Fourier space method
in order to investigate the sensitivity of our algorithm
to the various parameters,
but critical examples are demonstrated using
the full-covariance method.

Whenever we use the Fourier space method,
we find that the end-matching technique
improves the results (see \S\ref{sec:FitPars}), therefore, unless specified otherwise,
our frequency space solutions apply
end-matching to the light curve prior to processing.
The default parameters used in our simulations are listed in Table~\ref{tab:sim}.
\begin{table}
	\centering
	\caption{List of the default simulated parameters. Length is the number of points in our time series. For the unevenly spaced time series, see text for details.}
	\label{tab:sim}
	\begin{tabular}{ll} 
	    \hline
	    Parameter & Value \\
	    \hline
$\tau$           & 25\,day \\
$\gamma$         & 2       \\
$\alpha_{1}$     & 1     \\
$\alpha_{2}$     & $1/2$ \\
$\sigma_{F}/F$   & $0.03$ \\
$StD(F)/\langle{F}\rangle$ & 0.1-0.12 \\
Length           & 1000 \\
		\hline
	\end{tabular}
\end{table}

In \S\ref{sec:example}, we present an example based on a single simulated light curve.
Examples of fitting parameters to multiple simulations,
and the importance of the end-matching technique, is presented
in \S\ref{sec:FitPars}.
In \S\ref{sec:FitFullCov} we present the results based on our full-covariance method.
In \S\ref{sec:SensA2A1} we discuss the sensitivity of the full covariance method to $\alpha_{2}/\alpha_{1}$.
In \S\ref{sec:FalseAlarms}, we test the suitability of our method as a detector
(i.e., false alarm rate).
The effect of using the wrong $\gamma$
is discussed in \S\ref{sec:wronggamma},
and the sensitivity of our method to variations in the light curve
parameters is further investigated in \S\ref{sec:sensitivity}.
Finally, in \S\ref{sec:uneq}, we present some results
for unevenly spaced time series.

\subsection{A single example}
\label{sec:example}

\begin{figure}
\centerline{\includegraphics[width=8cm]{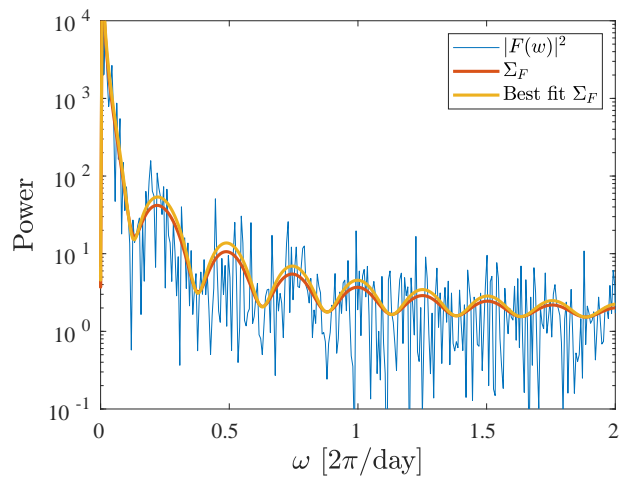}}
\caption{The power spectrum of a simulated light curve (blue line), as well as the actual power spectrum from which the light curve was generated (red line) and the best fitted $\Sigma_{F}$ (yellow line).
\label{fig:Flux_Sim11_PS_}}
\end{figure}

\begin{figure}
\centerline{\includegraphics[width=8cm]{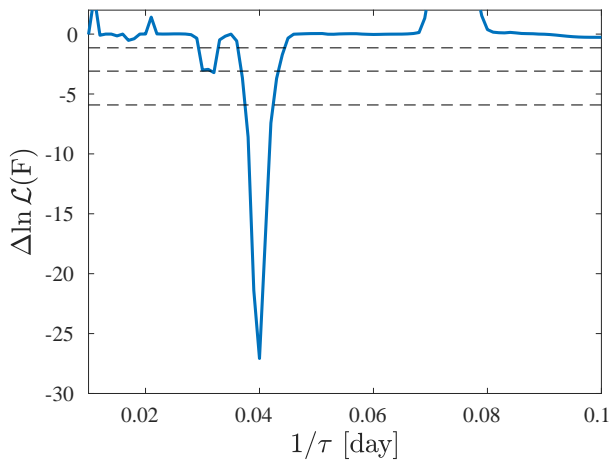}}
\caption{The $\Delta\ln\mathcal{L}$ as a function of $1/\tau$ for a single simulation using the Fourier space solution.
The $\Delta\ln\mathcal{L}$ is calculated, for each $\tau$, by subtracting the $-\ln\mathcal{L}$
of the null hypothesis (i.e., $\alpha_{2}=0$, $\tau=0$, but fitting $\gamma$)
from the minus log likelihood of the alternative hypothesis (i.e., fitting $\alpha_{1}$, $\alpha_{2}$, and $\gamma$). The horizontal dashed lines show the theoretical $1$, $2$, and $3\sigma$ confidence levels, assuming an $\chi^{2}/2$ distribution with two degrees of freedom.
This is calculated for the same simulated light curve as in
Figure~\ref{fig:Flux_Sim11_PS_}.
\label{fig:Flux_Sim11_PS_DL_Tau}}
\end{figure}

We start by presenting the results for a single simulation,
based on the parameters in Table~\ref{tab:sim}.
Figure~\ref{fig:Flux_Sim11_PS_} presents
the power spectrum of a simulated light curve, as well as the expectation
value of the power spectrum from which the light curve was generated, and the best fitted $\Sigma_{F}$ in the frequency domain (minimizing the minus of Equation~\ref{eq:logP_F_par}).
Next, for the same simulation, Figure~\ref{fig:Flux_Sim11_PS_DL_Tau}
shows the $\Delta\ln\mathcal{L}$ (frequency domain solution) 
as a function of $1/\tau$.
The $\Delta\ln\mathcal{L}$ is calculated, for each time-delay $\tau$, by subtracting the $-\ln\mathcal{L}$
of the null hypothesis ($H_{0}$; i.e., $\alpha_{2}=0$, $\tau=0$, but fitting $\gamma$)
from the minus log likelihood of the alternate hypothesis ($H_{1}$; i.e., fitting $\alpha_{1}$, $\alpha_{2}$, and $\gamma$).
Unless specified otherwise, the likelihood is calculated
for $1/\tau$ of $0.01$\,day$^{-1}$ to $0.1$\,day$^{-1}$ in steps of $10^{-3}$\,day$^{-1}$.

\begin{figure}
\centerline{\includegraphics[width=8cm]{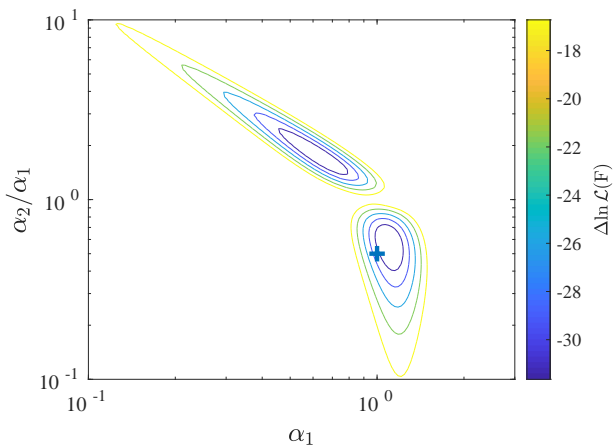}}
\caption{The log-likelihood surface as a function of $\alpha_{1}$ and $\alpha_{2}/\alpha_{1}$.
The plus mark shows the actual parameters used in the simulation.
The contours correspond to 1 to 5\,$\sigma$ assuming four degrees of freedom.
This is calculated for the same simulated light curve as in
Figure~\ref{fig:Flux_Sim11_PS_}, assuming the best fit $\tau$.
\label{fig:Flux_Sim11_a1_a1a2}}
\end{figure}

Our fit is done by minimizing $\Delta\ln\mathcal{L}$ over the relevant parameters (e.g., $\alpha_{i}$, $\gamma$) for every
$\tau$ separately. Therefore, in practice our hypotheses are not nested.
Nevertheless, we assume that there is a difference of two degrees of freedom between the two hypotheses, which is merely an approximation.
The horizontal dashed lines show the theoretical $1$, $2$, and $3\sigma$ confidence levels,
assuming an $\chi^{2}$ distribution with two degrees of freedom (where $\Delta{\chi^{2}}/2=\Delta{\ln\mathcal{L}}$).
It is clear that the actual time delay is recovered with high significance.
In Figure~\ref{fig:Flux_Sim11_a1_a1a2},
we present, for the same simulation and the best fit $\tau$, the log-likelihood surface
as a function of $\alpha_{1}$ and $\alpha_{2}/\alpha_{1}$.
The plus symbol shows the actual parameters used in the simulation.
The contours correspond to one to five $\sigma$ assuming four degrees of freedom,
while the color coding shows the $\Delta\ln\mathcal{L}$ compared to the null hypothesis.
The degeneracy between the positive and negative time delay and the $\alpha_{2}/\alpha_{1}$ below one and above one, is clearly seen.
We can also see (based on the $\Delta\ln\mathcal{L}$ values [see color bar]) that the
null hypothesis is rejected at a very high confidence level.

\subsection{Fitting the time delay}
\label{sec:FitPars}

\begin{figure}
\centerline{\includegraphics[width=8cm]{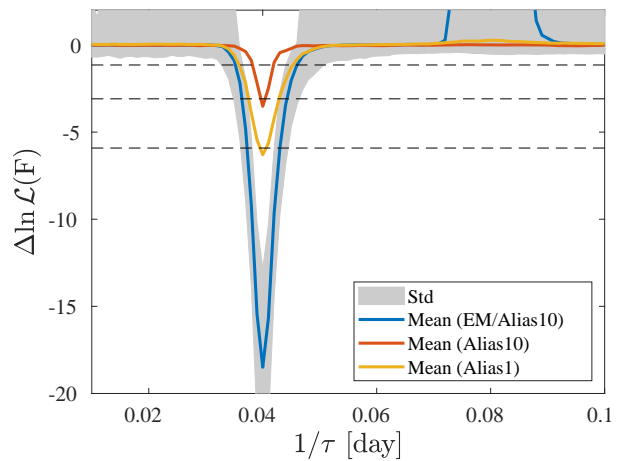}}
\caption{The $\Delta\ln\mathcal{L}$
as a function of $1/\tau$.
The solid lines show the mean of fitting 1000 simulated light curves,
for three cases: oversampling by a factor of 10 and end-matching (blue: EM/Alias10);
oversampling by a factor of 10 without end-matching (red; Alias10);
and no oversampling and no end-matching (yellow: Alias1).
The gray region shows the 1-$\sigma$ quantile of the EM/Alias10 simulations (blue line)
as a function of $1/\tau$.
The horizontal dashed lines show the theoretical 1, 2, and $3\sigma$ lines of rejecting
the null hypothesis, assuming two degrees of freedom.
\label{fig:Flux_Sim11_fitgamma_wEM_DL_Tau}}
\end{figure}

\begin{figure}
\centerline{\includegraphics[width=8cm]{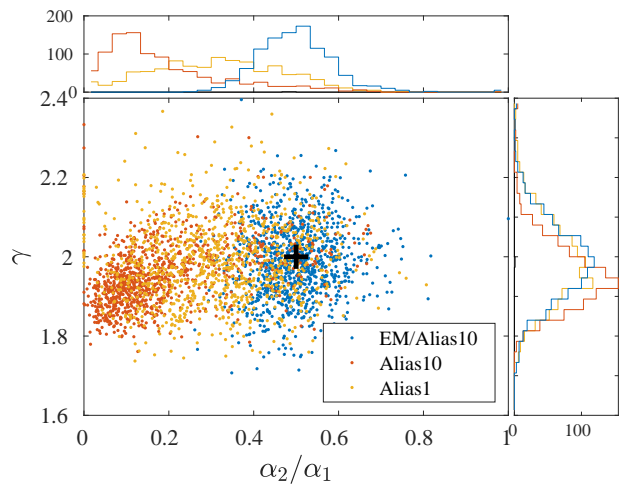}}
\caption{The distribution of the best-fitted $\alpha_{2}/\alpha_{1}$ and $\gamma$
parameters in the same simulations presented in Figure~\ref{fig:Flux_Sim11_fitgamma_wEM_DL_Tau},
and the same color coding.
The plus marker shows the actual value of the parameters used in the simulations.
\label{fig:Flux_Sim11_fitgamma_wEM_A2A1_gamma}}
\end{figure}

Next, we tested the ability of our algorithm
to identify the correct time delay and to fit the free parameters
using the Fourier space method.
Again, we use the simulation parameters in Table~\ref{tab:sim}
and for each time delay, we fit the free parameters
$\alpha_{1}$, $\alpha_{2}$, and $\gamma$.
Figure~\ref{fig:Flux_Sim11_fitgamma_wEM_DL_Tau} presents the $\Delta\ln\mathcal{L}$
as a function of $1/\tau$.
The solid lines show the mean $\Delta\ln\mathcal{L}$ of fitting 1000 simulated light curves,
for three cases: light curves that were generated using oversampling by a factor of 10 and end-matching applied before processing (blue: EM/Alias10);
oversampling by a factor of 10 without end-matching (red; Alias10);
and no oversampling and no end-matching (yellow: Alias1).
The gray region shows the 1-$\sigma$ quantile of the EM/Alias10 (blue line) simulations,
as a function of $1/\tau$.
The gray region deviates from the 1-$\sigma$ confidence level (upper dashed line). In fact the standard deviation of the $\Delta\ln\mathcal{L}$
of the simulations seems to underestimate the uncertainties and extend upward
more than naively expected.
This is presumably because ignoring the covariance results in under estimation of the noise in the null hypothesis.
In \S\ref{sec:FitFullCov} we further discuss this point and show that the full covariance method provides better results.
For the same simulations, Figure~\ref{fig:Flux_Sim11_fitgamma_wEM_A2A1_gamma}
shows the distribution of the best-fitted $\alpha_{2}/\alpha_{1}$ and $\gamma$
parameters.
These plots demonstrate that the end-matching procedure has an important
effect on the analysis in Fourier space, and it returns parameters that are less biased
compared to the case where we do~not apply end-matching.

\subsection{Fitting using the full covariance}
\label{sec:FitFullCov}

So far we have used the Fourier space solution.
Here we show that employing the full covariance method
is more powerful, and that the Fourier space method somewhat
underestimates the uncertainties.

\begin{figure}
\centerline{\includegraphics[width=8cm]{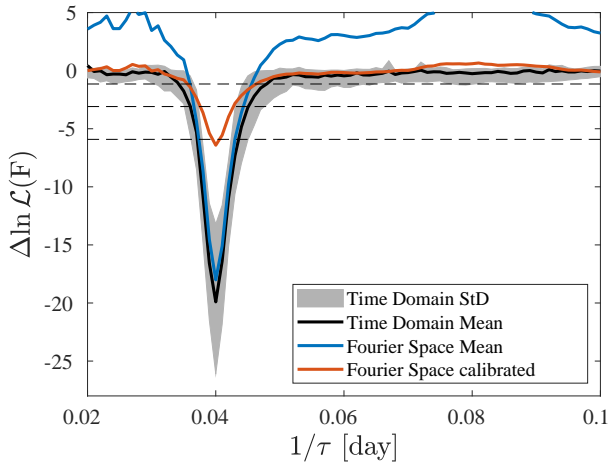}}
\caption{The $\Delta{\ln\mathcal{L}}$
as a function of $1/\tau$.
The black solid line shows the mean $\Delta{\ln\mathcal{L}}$ over 100 simulated light curves fitted with the full-covariance method, while the gray region shows the standard deviation of
the $\Delta{\ln\mathcal{L}}$ of these fits.
The blue line presents the mean over 100 simulations using the Fourier space fit, while the red line is the same after calibrating the blue line to its offset and noise (see text).
The dashed lines are like in Figure~\ref{fig:Flux_Sim11_fitgamma_wEM_DL_Tau}.
\label{fig:Flux_Sim11_TD}}
\end{figure}

We simulate light curves using the parameters in Table~\ref{tab:sim}.
Here, we do~not apply the end-matching process to the data prior to the fit.
We generated 100 simulated light curves, and for each light curve we fitted $\alpha_{1}$, $\alpha_{2}$, and $\tau$, assuming $\gamma$ is known.
In Figure~\ref{fig:Flux_Sim11_TD} we show the mean over all simulations (black solid line) of the
$\Delta\ln\mathcal{L}$
as a function of $1/\tau$.
The gray region shows the 1-$\sigma$ quantile of the $\Delta\ln\mathcal{L}$ over all the simulations.
The blue line shows, the same for an analysis in the Fourier space (without the full covariance).
At first glance, the results from the two methods look similar.
However, a careful comparison
between Figure~\ref{fig:Flux_Sim11_fitgamma_wEM_DL_Tau} and Figure~\ref{fig:Flux_Sim11_TD} shows that when the Fourier method is used (see Figure~\ref{fig:Flux_Sim11_fitgamma_wEM_DL_Tau}) the 1-$\sigma$ quantile is biased towards positive values, compared to what we get using the full covariance (Figure~\ref{fig:Flux_Sim11_TD}).
If we normalize the blue line, by subtracting its median, and dividing it by its relative standard deviation (compared to that of the black line), 
we get the red line, which has reduced sensitivity
compared to the full-covariance method.

A possible explanation for this behaviour is as follows.
In the first order approximation,
adding a constant to the full-covariance matrix will change the standard deviation (i.e., it will effect mainly $H_{0}$).
This point demonstrates the importance of using the full-covariance matrix,
and comparing it to the null hypothesis.

\subsection{Sensitivity to $\alpha_{2}/\alpha_{1}$}
\label{sec:SensA2A1}

\begin{figure}
\centerline{\includegraphics[width=8cm]{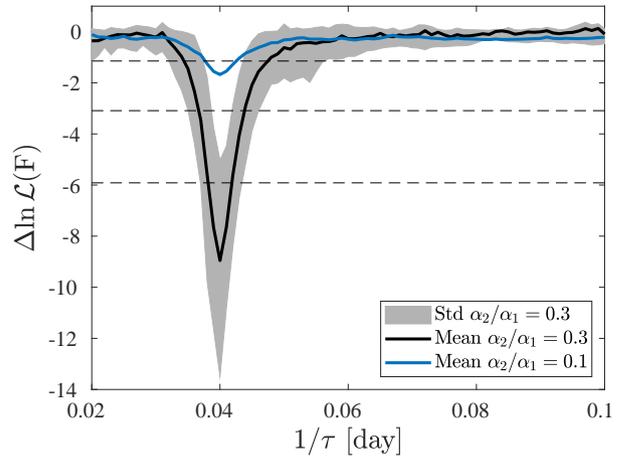}}
\caption{The $\Delta{\ln\mathcal{L}}$
as a function of $1/\tau$.
The solid line shows the mean $\Delta{\ln\mathcal{L}}$ over 100 simulated light curves fitted with the full-covariance method.
The black (blue) line is for $\alpha_{2}/\alpha_{1}=0.3$ ($0.1$).
The gray region shows the standard deviation of
the $\Delta{\ln\mathcal{L}}$ of the black line.
The dashed line are like in Figure~\ref{fig:Flux_Sim11_fitgamma_wEM_DL_Tau}.
\label{fig:Flux_Sim11_TD_A2}}
\end{figure}

Here we discuss the sensitivity of our algorithm
to $\alpha_{2}/\alpha_{1}$, when using the full-covariance method.
We generated 100 simulations using the parameters
in Table~\ref{tab:sim}, but this time with $\alpha_{2}/\alpha_{1}=0.3$,
and $\alpha_{2}/\alpha_{1}=0.1$.
In Figure~\ref{fig:Flux_Sim11_TD_A2} we present the, mean over 100 simulations, $\Delta{\ln\mathcal{L}}$
as a function of $1/\tau$ for $\alpha_{2}/\alpha_{1}=0.3$ (black line),
and $\alpha_{2}/\alpha_{1}=0.1$ (blue line).
The gray zone shows the 1-$\sigma$ quantile range of the 
$\alpha_{2}/\alpha_{1}=0.3$ simulations.
This plot suggests, that under reasonable assumptions, our method
may be able to detect systems with $\alpha_{2}/\alpha_{1}\gtorder0.1$.

\subsection{False alarms}
\label{sec:FalseAlarms}

In order to test our method as a detector, we would like to estimate the false alarm rate
assuming the null hypothesis is correct.
We test it using both the Fourier space method (1000 simulations),
and the time domain and full-covariance method (100 simulations).
We use the simulation parameters in Table~\ref{tab:sim},
but with $\alpha_{2}=0$ and $\tau=0$.
For each simulation, we attempt fitting
the flux-model, with known $\gamma$,
for $1/\tau$ in the range of $1/100$ to $1/10$, with steps of $1/1000$\,day$^{-1}$.

\begin{figure}
\centerline{\includegraphics[width=8cm]{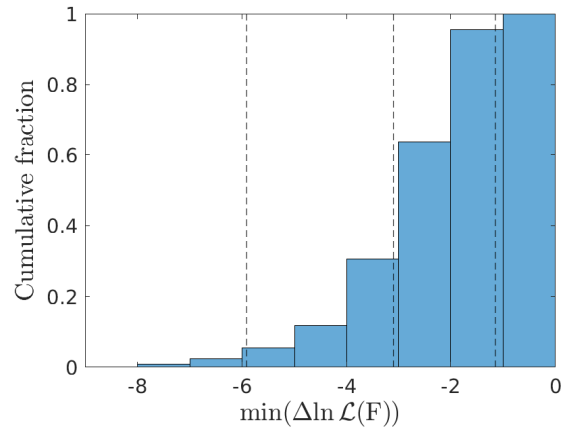}}
\caption{The cumulative histogram of the minimum $\Delta\ln\mathcal{L}$
in each $H_{0}$ simulation,
using the Fourier space fit.
The vertical dashed lines shows the 1, 2, and $3\sigma$ confidence levels for detection, assuming one time delay is tested and
two degrees of freedom.
\label{fig:Flux_Sim11_EM_Alias10_H0_hist}}
\end{figure}

\begin{figure}
\centerline{\includegraphics[width=8cm]{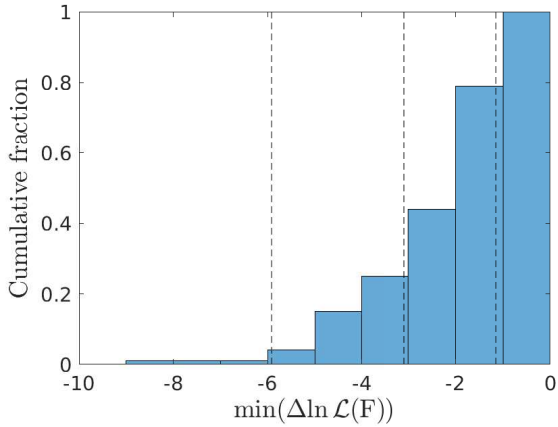}}
\caption{Like Figure~\ref{fig:Flux_Sim11_EM_Alias10_H0_hist},
but for the time delay and full-covariance method.
\label{fig:Flux_Sim11_TD_H0_hist}}
\end{figure}

The cumulative histogram of the minimum of $\Delta\ln\mathcal{L}$
in each simulation,
is shown in Figure~\ref{fig:Flux_Sim11_EM_Alias10_H0_hist}
for the Fourier space solution,
while Figure~\ref{fig:Flux_Sim11_TD_H0_hist} is for the full covariance method.
The vertical dashed lines show the 1, 2, and $3\sigma$,
assuming two degree of freedom.
The vertical lines assumes only one time delay is tested.
Since in each simulation we perform about 90 trials (for each time delay),
these lines underestimate the confidence level.
However, the trials are not completely independent,
and indeed correcting for the number of trials results
in a large overestimation of the confidence level.

\subsection{The effect of using the wrong $\gamma$}
\label{sec:wronggamma}

\begin{figure}
\centerline{\includegraphics[width=8cm]{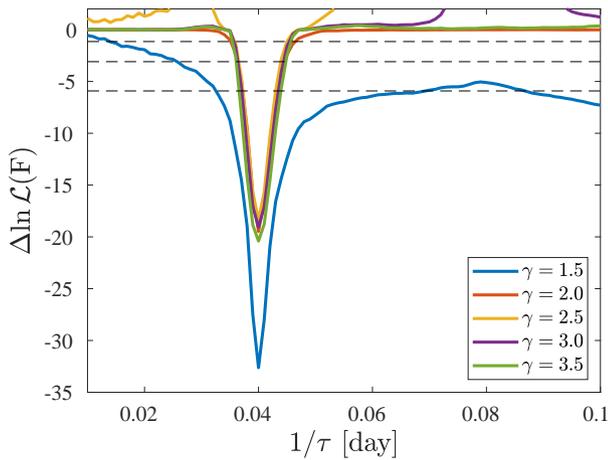}}
\caption{The $\Delta\ln\mathcal{L}$
as a function of $1/\tau$. The lines show the mean over 1000 simulated light curves.
In all the simulations, we used the light curve parameters in Table~\ref{tab:sim},
including $\gamma=2$, but in the fitting process we assumed that $\gamma$ is different
(as indicated in the legend).
\label{fig:Flux_Sim11_EM_Alias10_wronggamma_DL_Tau}}
\end{figure}

Since the power-law $\gamma$ of quasars is only roughly known,
we tested what happens if we use the simulation parameters in Table~\ref{tab:sim}
(with $\gamma=2$) but in the fit we assume the wrong $\gamma$.
Figure~\ref{fig:Flux_Sim11_EM_Alias10_wronggamma_DL_Tau} shows the $\Delta\ln\mathcal{L}$
as a function of $1/\tau$, calculated using the Fourier domain method. Each line is for the mean over 1000 simulations
and forcing a different value of $\gamma$, as indicated in the legend.
We conclude that our method can work even if we use the wrong assumption
on the spectral shape.

\subsection{Sensitivity to the various parameters}
\label{sec:sensitivity}

\begin{figure}
\centerline{\includegraphics[width=8cm]{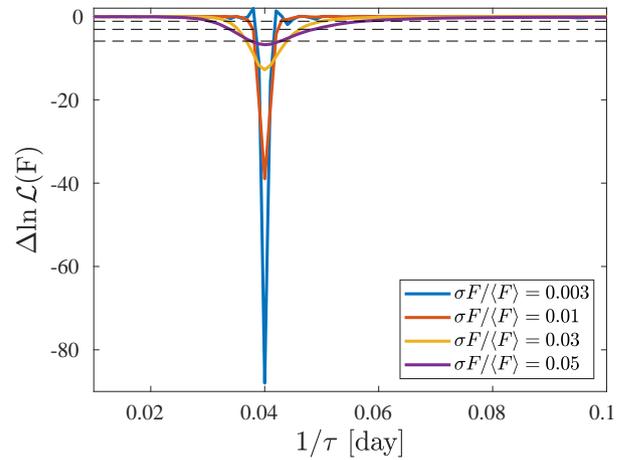}}
\caption{The mean over 1000 simulations of $\Delta\ln\mathcal{L}$ as a function of $1/\tau$
for the simulation parameters in Table~\ref{tab:sim} with several different values of $\sigma_{F}/F$
(values indicated in the legend).
\label{fig:Flux_Sim11_EM_Alias10_varsigmaf_DL_Tau}}
\end{figure}

\begin{figure}
\centerline{\includegraphics[width=8cm]{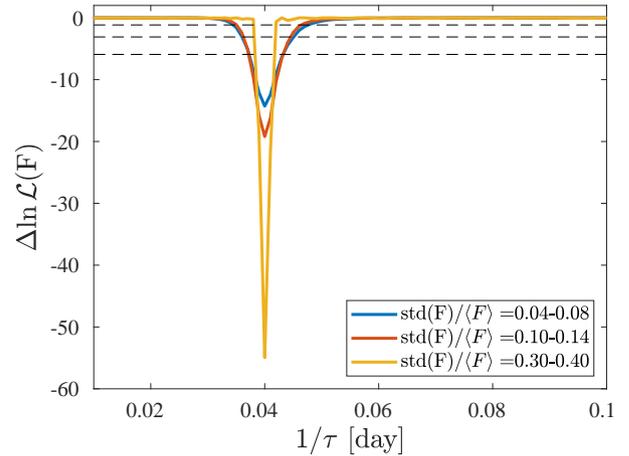}}
\caption{The same as Figure~\ref{fig:Flux_Sim11_EM_Alias10_varsigmaf_DL_Tau} but for different values of StD(F)$/\langle{F}\rangle$.
\label{fig:Flux_Sim11_EM_Alias10_varstd_DL_Tau}}
\end{figure}

\begin{figure}
\centerline{\includegraphics[width=8cm]{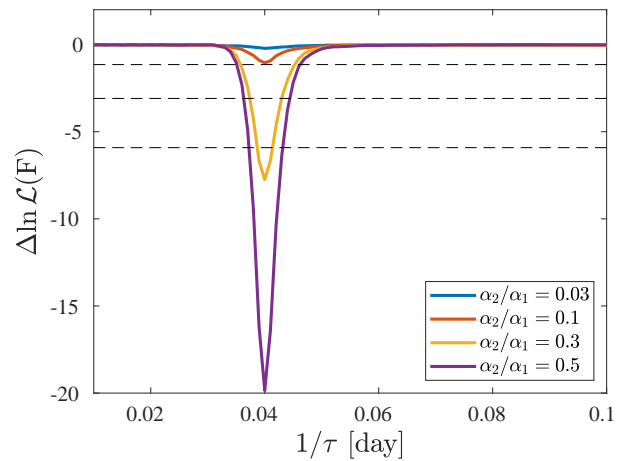}}
\caption{The same as Figure~\ref{fig:Flux_Sim11_EM_Alias10_varsigmaf_DL_Tau} but for different values of $\alpha_{2}$.
\label{fig:Flux_Sim11_EM_Alias10_vara2_DL_Tau}}
\end{figure}

\begin{figure}
\centerline{\includegraphics[width=8cm]{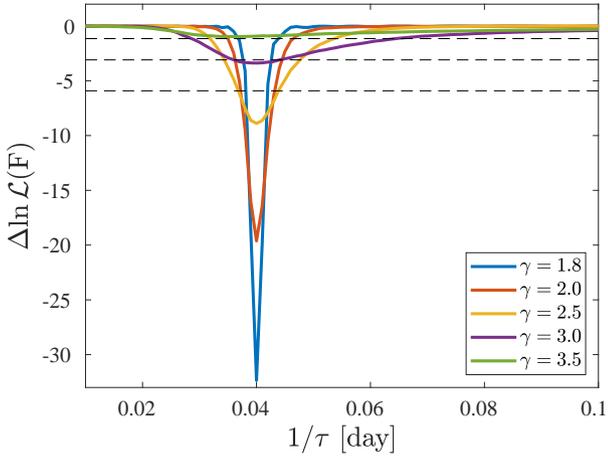}}
\caption{The same as Figure~\ref{fig:Flux_Sim11_EM_Alias10_varsigmaf_DL_Tau} but for different values of $\gamma$.
\label{fig:Flux_Sim11_EM_Alias10_vargamma_DL_Tau}}
\end{figure}

In this section, we present the performance of our method
when some of the parameters in Table~\ref{tab:sim} are modified.
In each set of simulations, we fix all the parameters to their values in Table~\ref{tab:sim}
and change only one parameter at a time.
Figures~\ref{fig:Flux_Sim11_EM_Alias10_varsigmaf_DL_Tau}--\ref{fig:Flux_Sim11_EM_Alias10_vargamma_DL_Tau}
show the mean over 1000 simulations of $\Delta\ln\mathcal{L}$ as a function of $1/\tau$ for the variation
of four parameters:
$\sigma_{F}/F$,
StD(F)$/\langle{F}\rangle$,
$\alpha_{2}$,
and $\gamma$, respectively.
The fits are done in Fourier space. For the case of   $\alpha_{2}/\alpha_{1}$ we also present a sensitivity test using the full covariance method in \S\ref{sec:FitFullCov}.

As expected, these plots demonstrate that our method gains sensitivity
when $\sigma_{F}/F$ decreases,
StD(F)$/\langle{F}\rangle$ increases,
$\alpha_{2}/\alpha_{1}$ increases,
and $\gamma$ decreases.
The reason that our method is more sensitive for smaller values of $\gamma$ is presumably because 
smaller values of $\gamma$ mean that the light curve is less correlated,
and hence effectively contains more independent data points.

\subsection{Unevenly spaced data}
\label{sec:uneq}

There are several ways to treat unevenly spaced data.
Simplistic methods include interpolating the data into an evenly spaced grid.
The most accurate method is to use our full covariance algorithm.
The advantage of this method is that no interpolation is required,
and that the accurate model-correlations between observations
are used.
The disadvantage of this method is that the integral in Equation~\ref{eq:SigmaT_t} needed to be evaluated for $\approx N^{2}/2$
time differences, instead of $N$ time differences in the evenly spaced case.
However, these integrals can be calculated on a grid in advance and
we can use interpolation to evaluate them.

\begin{figure}
\centerline{\includegraphics[width=8cm]{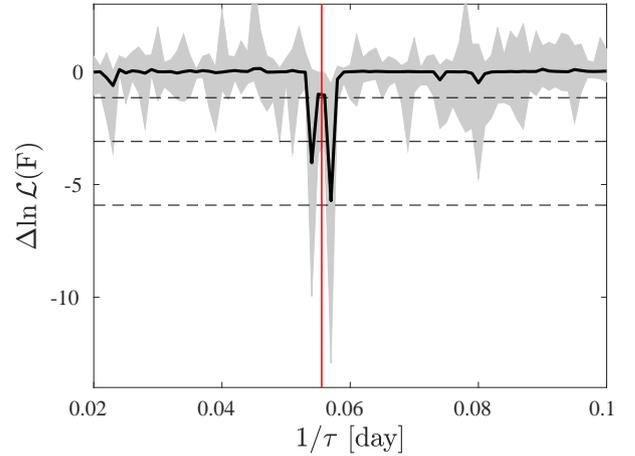}}
\caption{The median over 30 simulations of $\Delta\ln\mathcal{L}$ as a function of $1/\tau$
for the simulation of the unevenly spaced light curves using the full covariance method.
The gray patch represents the 1-$\sigma$ quantile of the simulations,
while the dashed lines are like in Figure~\ref{fig:Flux_Sim11_PS_DL_Tau}.
The vertical red line shows the simulated time delay of 18\,days.
\label{fig:Flux_Sim11_Tau18_unevenly_TD}}
\end{figure}

To demonstrate the method on unevenly spaced data,
we generate a three year quasar light curve using the parameters in Table~\ref{tab:sim}, but a time delay of 18\,days,
and time sampling of $0.1$\,day.
Next, we assume a survey cadence of 1\,day, with 0.05\,day random shifts in time,
and we remove about 20\% of the points in the same phase on each synodic month cycle.
We also select data points over 270\,days in annual cycles.
We interpolate the evenly spaced simulated light curve onto the unevenly spaced grid
and apply our full covariance method.
Figure~\ref{fig:Flux_Sim11_Tau18_unevenly_TD} presents the median $\Delta\ln\mathcal{L}$ 
as a function of $1/\tau$, over 30 simulations.
There are two minima just around the simulated time delay of 18\,days.
These are presumably due to aliasing of the true time delay with the window function of the data.

\section{Code}
\label{sec:code}

We implemented our algorithm in two environments:
a Python code, and a MATLAB package.
The two packages are implemented in slightly different ways.

The Python code, available from GitHub\footnote{https://github.com/ofersp/qtd}, provides a reference code with most of the necessary functions, simulations and tests,
but it does~not include a flexible fitting procedure
and the implementation of the full-covariance method.
These extra capabilities are planed to be added soon.

The MATLAB version is available from GitHub\footnote{https://github.com/EranOfek/TimeDelay}.
The package include the code used both in this paper and in Paper~II.
This package includes functions to calculate the likelihood,
fit the parameters, generate simulated light curves,
and to analyze unevenly spaced data sets.

\section{Discussion}
\label{sec:disc}

We present a novel method for assessing whether a light curve
is composed of two or more time-delayed light curves, and if so,
for measuring the time delay between the components.
This algorithm can be used only if a rough statistical model from which the light curve is generated is known
and if such a model contains power over a wide range of frequencies.
Such a method has the potential to measure reverberation time delays,
search for lensed quasars and measure their time delays,
and possibly measure
time delays in millimeter-wavelength data of unresolved observations of the silhouette
of super-massive black holes in galactic centers (e.g., \citealt{Hadar+2020_BlackHole_PhotonRingAutoCorrelation}).
Furthermore,
we comment on how to fit a power-law model to a power spectrum.
Specifically, we argue that the commonly used $\chi^{2}$ statistics is not adequate to this problem.
This is because the probability distribution of the power-spectrum
around its expectation value is not Gaussian.
Furthermore, our methods can be adapted to fit a power-law to unevenly spaced data,
or to other problems that involves shifts
(e.g., Speckle imaging; will be discussed in a separate publication).

In a companion paper (Springer \& Ofek 2020; Paper~II)
we present a more elaborate method that uses the total flux and the center-of-light position
of the system for searching for lensed quasars and measuring their time-delay.
The center-of-light position contains information on the flux ratio between the images.
In turn, the flux ratio along with the combined flux can be used to reconstruct the source light curve.
If such astrometric data contains information and it is available,
the method described in Paper~II is more powerful.

With the growing amount of data from previous, existing and upcoming sky surveys (e.g., \citealt{Tyson+2001_LSST}; \citealt{Law+2009_PTF};
\citealt{Tonry2011_ATLAS_SurveyCapability};
\citealt{Chambers+2016_PS1_Surveys}; \citealt{Bellm+2019_ZTF_Overview};
\citealt{Ofek+BenAmi2020_Grasp_SkySurvrys_CostEffectivness}),
these methods have the potential to uncover a significant fraction of all lensed quasars,
and possibly even measure the reverberation time delays for a large sample of quasars.

We provide Python and MATLAB implementations and
we test our methods on simulated data.
In a future work we will attempt to apply these methods to real data.
Our simulations suggest that applying the flux-only method to real data will likely require
at least hundreds of photometric data points per source,
with a photometric precision better than 5\%, and $\alpha_{2}/\alpha_{1}\gtorder0.1$.
As usual, applying these methods to real data will likely require more efforts
to deal with additional complexities.

\section*{Acknowledgements}

We thank Oren Raz, Barak Zackay, Brad Cenko, and Ariel Goobar for many enlightening discussions.
E.O.O. is grateful for the support of
grants from the 
Willner Family Leadership Institute,
Ilan Gluzman (Secaucus NJ), Madame Olga Klein - Astrachan,
Minerva foundation,
Israel Science Foundation,
BSF, BSF-transformative, Israel Ministry of Science,
Weizmann-Yale, and Weizmann-UK.

\section*{Data Availability}

The scripts used to simulate and test the method are available via Github as indicated in the main text.

\bibliographystyle{mnras}
\bibliography{papers.bib}


\newpage
\onecolumn

\appendix

\section{Some properties of a real multivariate normal distribution}
\label{App:cmvn}

A real valued random vector $x \in \mathbb{R}^n$ is said to be distributed as a multivariate normal (MVN) if it has the following probability density function (PDF):
\begin{equation}
P(x|\mu_x, \Sigma_x) = \frac{e^{-\frac{1}{2}(x-\mu_x)^T\Sigma_x^{-1}(x-\mu_x)}}{\sqrt{\det{(2\pi\Sigma_x)}}},
\end{equation}
where $\det{M}$ denotes the determinant of matrix $M$ and the $T$-sign denotes the matrix transpose. The MVN distribution is characterized by the following mean vector and covariance matrix:
\begin{eqnarray}
\mu_x &\equiv& \operatorname{E}\left[x\right], \\
\Sigma_x &\equiv& \operatorname{E}\left[(x-\mu_x)(x-\mu_x)^T\right].
\end{eqnarray}
We denote that the random vector $x$ has the above distribution by writing $x \sim N(\mu_x, \Sigma_x)$.

\section{Generalization to the multi-image case}
\label{App:multi_image}

When more than two images of a quasar are present, some of the formulae of \S\ref{sec:method} need to be generalized to accommodate this. We start by updating Equation \ref{eq:F}  of the time domain's total observed flux. Assuming the combined light curve is composed of $n$ source light curves, each having a flux factor $\alpha_i$ and time-delays $\tau_i$ relative to image 1, then
\begin{eqnarray}
\label{eq:flux_multi}
F(t) &=& \phi(t) + \epsilon_F(t) = \alpha_0 + \sum_{i=1}^n \alpha_i f(t+\tau_i) + \epsilon_f(t), \\
\end{eqnarray}
and Equations \ref{eq:Phi_of_omega} and \ref{eq:F_of_omega} for the noiseless and noisy flux in the frequency domain become
\begin{eqnarray}
    \widehat{\phi}(\omega) &=& \alpha_0 \delta(\omega) + \left(\sum_{i=1}^n \alpha_i e^{i\omega\tau_i}\right) \widehat{f}(\omega), \\
    \widehat{F}(\omega) &=& \widehat{\phi}(\omega) + \widehat{\epsilon}(\omega).
\end{eqnarray}
This allows us to update the expression,
in Equation~\ref{eq:Sigma_phi},
for the variance of the noiseless total flux:
\begin{eqnarray}
\label{eq:phi_var_multi}
    \operatorname{E}\left[\widehat{\phi}(\omega)\widehat{\phi}^*(\omega)\right] \equiv \Sigma_\phi(\omega) &=& 
    \operatorname{E}\left[
    \left(\sum_{i=1}^n \alpha_i e^{i\omega\tau_i}\right)
    \widehat{f}(\omega)\widehat{f}^*(\omega)
    \left(\sum_{i=1}^n \alpha_i e^{-i\omega\tau_i}\right)
    \right] \\ \nonumber
    &=& \frac{1}{|\omega|^\gamma}
    \left[
    \sum_{i=1}^n \alpha_i^2 +
    2\sum_{i>j}\alpha_i\alpha_j \cos(\omega[\tau_i-\tau_j])
    \right],
\end{eqnarray}
which is valid for non-zero frequencies.
By redefining $\Sigma_F(\omega) \equiv \Sigma_\phi(\omega) + \widehat{\sigma}_F^2$ we can also update Equation~\ref{eq:log_P_flux} for the log-likelihood of observing the total flux given the model parameters
\begin{eqnarray}
\label{eq:log_P_flux_multi}
    \log P(\widehat{F} | \tau_i, \alpha_i) = -\frac{1}{2}\ln \det{(2\pi\Sigma_{F})} - 
    \frac{|\widehat{F}(\omega)|^2}{2\Sigma_F(\omega)}.
\end{eqnarray}

\section{Calculating log of the determinant}
\label{sec:logdet}

The log of the determinant of a matrix is calculated using the following algorithm.
We perform LU decomposition with partial factorization.
For input matrix $A$, LU decomposition returns $P$, $L$ and $U$ such that $PA = LU$, where
$L$ is a lower triangular matrix, and $U$ is an upper triangular matrix.
Next we calculate
\begin{eqnarray}
    C=\det(P) \prod{{\rm sign}({\rm diag}(U))},
\end{eqnarray}
where ${\rm sign}$ is the sign function and ${\rm diag}$ is the diagonal function.
and finally
\begin{equation}
    \ln{\det{A}} = \ln{C} + \sum{\ln{|{{\rm diag}{U}}|}}
\end{equation}

\section{The auto-correlation integral}
\label{sec:AutoCorr}

The integral in Equation~\ref{eq:SigmaT_t} diverges. Here we present two methods to solve this problem. The first method is to start the integration from a small positive frequency $\omega_{0}$.
The true $\omega_{0}$ from which the data was generated is, in most cases, unknown.
However, we show that the results are not very sensitive to the chosen $\omega_{0}$.
The second method is to rewrite the diverging integral as a sum of converging and diverging integrals, where the diverging integral depends only on $\gamma$,
and we set the diverging integral to zero.
We show that the two approaches give similar results.

We can rewrite $\Sigma_{T}(t_{i},t_{j})$ as
\begin{equation}
    \Sigma_{T}(t_{i},t_{j}) = 2(\alpha_{1}^{2}+\alpha_{2}^{2})(G_{0}+G_{1}) + 4\alpha_{1}\alpha_{2}(G_{0}+G_{1}'+G_{2}) + \delta(t_{j}-t_{k})\sigma_{F}^{2},  
    \label{eq:SigmaT}
\end{equation}
where
\begin{equation}
    G_{0}(\gamma) = \int_{\omega_{0}}^{\infty}{\omega^{-\gamma}   d\omega} = \frac{\omega_{0}^{(1-\gamma)}}{\gamma-1},
    \label{eq:G0}
\end{equation}
\begin{equation}
    G_{1}(\Delta{t},\gamma) = \int_{\omega_{0}}^{\infty}{\frac{\cos{(\omega\Delta{t})}-1}{\omega^{\gamma}}   d\omega},
    \label{eq:G1}
\end{equation}
\begin{equation}
    G_{1}'(\tau,\gamma) = \int_{\omega_{0}}^{\infty}{\frac{\cos{(\omega\tau)}-1}{\omega^{\gamma}}   d\omega},
    \label{eq:G1}
\end{equation}
and
\begin{equation}
    G_{2}(\Delta{t},\tau,\gamma) = \int_{\omega_{0}}^{\infty}{\frac{(\cos{(\omega\Delta{t})}-1)\cos{\omega\tau}}{\omega^{\gamma}}   d\omega}.
    \label{eq:G2}
\end{equation}
In Equation~\ref{eq:G0}, the analytic expression of the integral is correct only for $\gamma>1$.
This representation of Equation~\ref{eq:SigmaT_t} is important
because it makes the integral we need to calculate to be independent
of $\alpha_{1}$ and $\alpha_{2}$. Therefore, the number of integral evaluations
is reduced dramatically.
Next, by using L'Hôpital's rule we can
see that, for $\omega_{0}=0$, the integrals for $G_{1}$ and $G_{2}$ converge for $1<\gamma<3$.
This range is likely relevant for quasars (but maybe not for all quasars).
In practice, we calculate these integrals using numerical integration.
However, the integral for $G_{0}$,
where $\omega_{0}=0$,
diverges for any value of $\gamma$. An important fact is that
the integral for $G_{0}$
depends only on $\gamma$.

Our code supports two methods to calculate Equation~\ref{eq:SigmaT}.
The first is by setting $\omega_{0}$ to a small number.
If $\omega_{0}$ is unknown from the data, we recommend setting it to the lowest frequency of the observations.
The second method is to perform the integrals from zero to infinity,
but to set the integral of $G_{0}$ (Eq.~\ref{eq:G0}) to zero.
In cases, in which $\gamma$ is not a free parameter, $G_{0}$ is
a constant related (in first order) to the mean of the time series ($\mu_{F}$).
The examples in \S\ref{sec:simulations} are calculated using
the $G_{0}=0$ approach.

\begin{figure}
\centerline{\includegraphics[width=8cm]{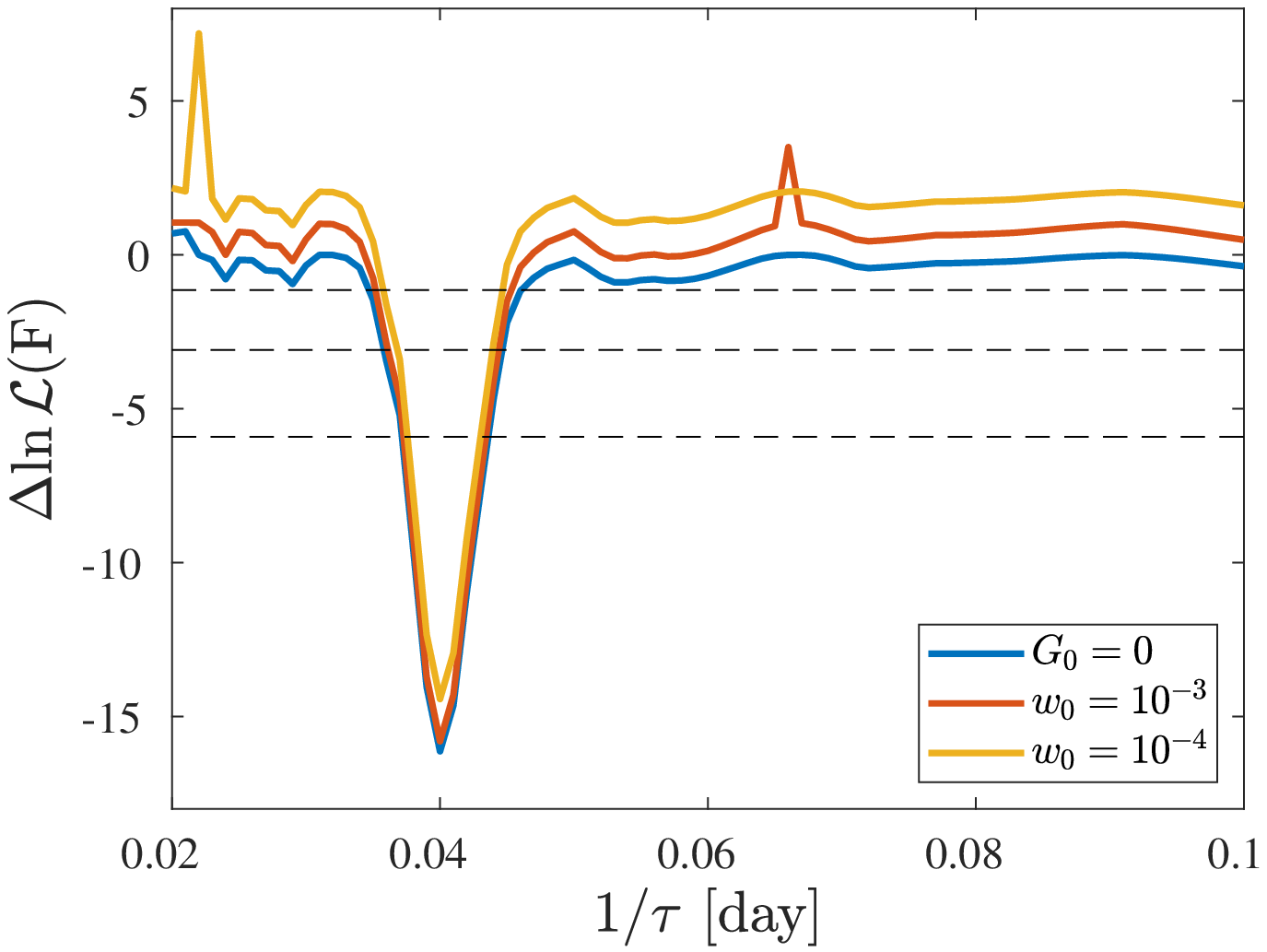}}
\caption{The $\Delta\ln\mathcal{L}$ as a function of $1/\tau$ for a single simulation using the parameters in Table~\ref{tab:sim},
but calculated using two different methods to calculate $\Sigma_{T}(t_{j}-t_{k})$.
In blue, the integral in Equation~\ref{eq:SigmaT} is calculated by setting $G_{0}=0$ and calculating $G_{1}$, $G_{1}'$, and $G_{2}$ between zero and infinity.
The other lines are based on integration of all the integrals (Eqs.~\ref{eq:G0}-\ref{eq:G2}) between $\omega_{0}$ to infinity,
where $\omega_{0}$ is $10^{-3}$ (red line) and $10^{-4}$ (yellow line).
The $\Delta\ln\mathcal{L}$ is calculated, for each $\tau$, by subtracting the $-\ln\mathcal{L}$
of the null hypothesis (i.e., $\alpha_{2}=0$, $\tau=0$)
from the minus log likelihood of the alternate hypothesis (i.e., fitting $\alpha_{1}$, $\alpha_{2}$). The horizontal dashed lines show the theoretical $1$, $2$, and $3\sigma$ confidence levels, assuming an $\chi^{2}/2$ distribution with two degrees of freedom.
\label{fig:Flux_Sim11_G0_treatment}}
\end{figure}

In order to demonstrate that the two approaches work well,
in Figure~\ref{fig:Flux_Sim11_G0_treatment} we present
the fitted $\Delta\ln\mathcal{L}$ as a function of $1/\tau$
for a single simulation, based on the two approaches,
and different values of $\omega_{0}$.
It is clear that the differences are small, and the two approaches work well.

\section{Light curve simulations}
\label{app:cycsim}

Here we describe the prescription we used to generate simulated light curves
(see also \citealt{Timmer+Koenig1995_GeneratingPowerLawNoise}).
Given that we are using the Fast Fourier Transform,
our method generate simulations that are evenly spaced and with
cyclic boundary conditions.

In order to extend these simulations to non-cyclic cases, we repeat the prescription for a time series which is twice as long as needed, and trim the final light curve by a factor of two.
In order to generate unevenly spaced data, we use short time steps and interpolate the time series into an unevenly spaced grid.

The method includes the following steps:

\begin{itemize}

\item First, a frequency realization of the original quasar flux, $\widehat{f}(\omega)$, is pseudo-randomly drawn to have a red-power spectrum at a set of discrete angular frequencies $\omega \in (-\frac{n}{2}+1, ..., 0, \frac{n}{2}-1)\frac{2\pi}{T}$, where $n = T/\Delta t$. Here $T$ is the total time of the simulation, $\Delta t$ is the time interval between consecutive observations and $n$ is the number of temporal samples. To produce a red-power spectrum, each frequency component's real and imaginary parts are randomly drawn to have a uniform phase around the full unit circle and a normally distributed magnitude with mean zero and standard deviation of $|\omega|^{-\gamma/2}$. Additionally, the constant flux level $n f_{DC}$ is added to $\widehat{f}(0)$, where the prefactor $n$ is due to the definition of the discrete Fourier transform (DFT) used. To ensure the resulting temporal realization $f(t)$ has no imaginary components, negative frequency components of $\widehat{f}(\omega)$ are set to be complex conjugate symmetric to their matching positive components.

\item The resulting frequency domain noiseless total flux is set (according to Equation \ref{eq:Phi_of_omega}) to
$\widehat{\phi}(\omega) = (\alpha_1+\alpha_2 e^{i\omega\tau})\widehat{f}(\omega)$ and additionally the constant lensing galaxy flux $n\alpha_0$ is added to the zero frequency component $\widehat{\phi}(0)$.

\item In a similar fashion to the generation of the red noise, Gaussian noise $\widehat{\epsilon}_F(\omega)$ are generated in the frequency domain that would produce (following an inverse DFT) a real temporal flux noise $\epsilon_F(t)$ with a standard deviation $\sigma_F$.
The observational noise is then added to $\widehat{\phi}(\omega)$ to produce $\widehat{F}(\omega)$.

\item The $n$ temporal samples of $f(t)$, $f_1(t)$, $f_2(t)$, $\phi(t)$ and $F(t)$ are computed (using an inverse DFT) from their frequency realizations $\widehat{f}(\omega)$, $\alpha_1\widehat{f}(\omega)$, $\alpha_2 e^{i\omega\tau}\widehat{f}(\omega)$, $\widehat{\phi}(\omega)$ and $\widehat{F}(\omega)$, respectively.

\item The time series generated so far are evenly spaced. Although one can generate unevenly spaced time series in a similar fashion to the steps above, we choose to generate evenly spaced time series and re-sample them using interpolation, into unevenly spaced series. In the case of the red power-spectrum, where high frequency amplitudes are comparable to the noise, this scheme is justified.

\end{itemize}

\bsp	
\label{lastpage}

\end{document}